\documentclass[letter,11pt]{article}
\usepackage{jheppub} 

\usepackage[T1]{fontenc} 
\usepackage{slashed}
\usepackage{epsfig}
\usepackage{comment}
\usepackage{cancel}
\usepackage{bbm}
\usepackage{array}
\usepackage{bigints}
\usepackage{booktabs}
\usepackage{color}
\usepackage{dsfont}
\usepackage{float}
\usepackage{framed}
\usepackage{graphicx}
\usepackage{indentfirst}
\usepackage{mathrsfs}
\usepackage{multirow}
\usepackage{subdepth}
\usepackage{titlesec}
\usepackage[dotinlabels]{titletoc}
\usepackage{wrapfig}
\usepackage[all]{xy}
\usepackage[vcentermath]{youngtab}
\usepackage{relsize}
\usepackage{hyperref}
\numberwithin{equation}{section}

\usepackage[utf8]{inputenc}
\usepackage{slashed}
\usepackage{amsmath}
\usepackage{amsfonts}
\usepackage{amssymb}
\usepackage{mathtools}

\usepackage{cleveref}
\crefname{section}{§}{§§}
\Crefname{section}{§}{§§}

\def\e{{\epsilon}}

 \def\p{\partial}

\def\0{{(0)}}
\def\1{{(1)}}
\def\2{{(2)}}

\def\ci{{\mathscr I}}

\def\<{\langle }
\def\>{\rangle }

\def\G{{\Gamma}}
\def\cc{{\text{c.c.}}}
\def\o{{\omega}}

\newcommand{\bea}{\begin{eqnarray}}
\newcommand{\eea}{\end{eqnarray}}
\newcommand{\be}{\begin{equation}}
\newcommand{\ee}{\end{equation}}
\newcommand{\ba}{\begin{align}}
\newcommand{\ea}{\end{align}}

\newcommand{\w}[1]{\mbox{$\W_\infty[#1]$}}

   \makeatletter
  \let\over=\@@over \let\overwithdelims=\@@overwithdelims
  \let\atop=\@@atop \let\atopwithdelims=\@@atopwithdelims
  \let\above=\@@above \let\abovewithdelims=\@@abovewithdelims
\renewcommand\section{\@startsection {section}{1}{\z@}%
                                   {-3.5ex \@plus -1ex \@minus -.2ex}
                                   {2.3ex \@plus.2ex}%
                                   {\normalfont\large\bfseries}}

\renewcommand\subsection{\@startsection{subsection}{2}{\z@}%
                                     {-3.25ex\@plus -1ex \@minus -.2ex}%
                                     {1.5ex \@plus .2ex}%
                                     {\normalfont\bfseries}}

\newcommand{\pd}[2]{\frac{\partial #1}{\partial #2}}

\newcommand{\beq}{\begin{equation}}
\newcommand{\eeq}{\end{equation}}
\newcommand{\beqa}{\begin{eqnarray}}
\newcommand{\eeqa}{\end{eqnarray}}
\newcommand{\beqar}{\begin{eqnarray*}}

\newcommand{\ve}{{\varepsilon}}

\def\[{\big[}
\def\]{\big]}
\def\la{\langle}
\def\ra{\rangle}

\def\mjj{{\mathbb J}}
\def\mqq{{\mathbb Q}}

\def\e{{\epsilon}}
\def\ve{{\varepsilon}}

\def\g{{\gamma}}
\def\o{{\omega}}
\def\a{{\alpha}}
\def\b{{\beta}}

\def\d{{\delta}}

\def\w{\omega}

\def\be{{\bar \epsilon}}

\def\CA{{\mathcal A}}
\def\CB{{\mathcal B}}

\def\CI{{\mathcal I}}
\def\ci{{\mathscr I}}
\def\CJ{{\mathcal J}}
\def\CK{{\mathcal K}}

\def\CL{{\mathcal L}}

\def\CO{{\mathcal O}}

\def\CQ{{\mathcal Q}}
\def\CR{{\mathcal R}}
\def\CS{{\mathcal S}}

\def\CY{{\mathcal Y}}

\def\SO{{\mathscr O}}

\newcommand{\bra}[1]{\langle\,   #1\,    |}
\newcommand{\ket}[1]{ |\,   #1 \,  \rangle}
\newcommand{\braket}[2]{\langle\,   #1 \, | \, #2 \, \rangle}

\def\mrr{{\mathbb R}}

\def\mqq{{\mathbb Q}}

\def\pmm{{(\pm)}}

\def\+{{(+)}}
\def\-{{(-)}}
\def\0{{(0)}}
\def\1{{(1)}}
\def\2{{(2)}}
\def\3{{(3)}}

\def\d{{(d)}}

\title{\boldmath New Magnetic Symmetries in $(d+2)$-Dimensional QED}

\author[]{Temple He$^1$}
\author[]{and Prahar Mitra$^2$}

\affiliation[]{$^1$Center for Quantum Mathematics and Physics (QMAP), University of California, Davis, CA 95616, USA}\affiliation[]{$^2$School of Natural Sciences, Institute for Advanced Study, Princeton, NJ 08540, USA}

\emailAdd{tmhe@ucdavis.edu}
\emailAdd{prahar21@ias.edu}

\abstract{Previous analyses of asymptotic symmetries in QED have shown that the subleading soft photon theorem implies a Ward identity corresponding to a charge generating divergent large gauge transformations on the asymptotic states at null infinity. In this work, we demonstrate that the subleading soft photon theorem is equivalent to a more general Ward identity. The charge corresponding to this Ward identity can be decomposed into an electric piece and a magnetic piece. The electric piece generates the Ward identity that was previously studied, but the magnetic piece is novel, and implies the existence of an additional asymptotic ``magnetic'' symmetry in QED.}

\begin{document} 
\maketitle
\flushbottom

\section{Introduction}

In recent years, an intricate relationship between soft theorems and asymptotic symmetries in asymptotically flat spacetimes has been discovered and extensively studied (for a detailed review of this subject, see \cite{Strominger:2017zoo}). It began with the discovery that the leading soft theorems in both four dimensional gauge and gravity theories are equivalent to Ward identities associated to charges generating asymptotic symmetries of the theory \cite{He:2014laa,He:2014cra,He:2015zea,Campiglia:2015qka,Kapec:2015ena,Dumitrescu:2015fej}. These results were later extended to all higher dimensions \cite{Kapec:2014zla,Kapec:2015vwa,He:2019jjk,Henneaux:2019yqq}.

The relationship between asymptotic symmetries and soft theorems became more intriguing when it was observed that there is also a relationship between the subleading soft theorems and asymptotic charges generating divergent large gauge symmetries in all dimensions \cite{Schwab:2014xua,Kapec:2014opa,Lysov:2014csa,Campiglia:2016hvg,Conde:2016csj,Laddha:2017vfh,Hirai:2018ijc,He:2019pll}. However, unlike the case involving the leading soft theorems, the subleading soft theorem is oftentimes a \emph{stronger} condition than the associated Ward identity. While the subleading soft theorem implies the Ward identity, the reverse is not necessarily true. 

Traditionally, one conjectures a `matching condition' that relates a specific component of a field at past null infinity $\ci^-$ to that at future null infinity $\ci^+$ (see Section \ref{sec:matching}),\footnote{These matching conditions were rigorously studied (and proved) in QED and gravity in \cite{Prabhu:2018gzs,Prabhu:2019fsp}.} which can then be massaged into a Ward identity (in the semi-classical theory) for the $S$-matrix. For example, in gauge theories one imposes such a matching condition for the radial electric field $E_r$, and the corresponding Ward identity is equivalent to the leading soft photon theorem. Similarly, in gravitational theories the matching condition for the electric part of the Weyl tensor is equivalent to the leading soft graviton theorem. Crucially, each independent matching condition leads to an independent Ward identity or soft theorem. This does not imply a contradiction for the $d$ leading soft photon theorems (one for each polarization), which are not all independent. Rather, they satisfy a trivial identity that results in a single independent leading soft theorem \cite{He:2019jjk}; this is the soft theorem that is equivalent to the matching condition described above. A similar argument holds for the leading soft graviton theorem as well. 

However, such an identity does not hold for the subleading soft photon theorem, thereby implying there are indeed $d$ independent subleading soft theorems. It is therefore not possible to demonstrate its equivalence with a Ward identity arising from matching the radial electric field. Rather, the matching condition for the radial electric field leads to a particular linear combination of the subleading soft theorem \cite{He:2019pll}, which we shall henceforth call the subleading \emph{electric} Ward identity. The origin (from the perspective of asymptotic symmetries or matching conditions) of the remaining $d-1$ independent soft theorems is, so far, unknown.

In this paper, we conjecture a matching condition for the $d$ angular components of the magnetic field (a vector matching condition), and then show that it is equivalent to the $d$ independent subleading soft photon theorems. Naturally, our ansatz that there are $d$ matching conditions instead of just one implies that the associated Ward identities must correspond to new symmetries. As it turns out, the subleading electric Ward identity corresponds precisely to one of the $d$ matching conditions. The remaining $d-1$ matching conditions then give rise to Ward identities corresponding to charges that generate magnetic large gauge transformations, and we shall call these Ward identities the subleading \emph{magnetic} Ward identities. This suggests that even though there are no global magnetic charges in our theory, there exists finite large gauge symmetries that are generated by asymptotic magnetic charges.

This paper is organized as follows. In Section \ref{sec:prelim}, we will summarize all the notations and conventions used throughout the paper. We will also derive the asymptotic expansion of the gauge field near $\ci^\pm$; because much of the technology used was introduced in \cite{He:2019jjk}, we refer the reader there for more details. In Section \ref{sec:ward}, we will conjecture the set of $d$ matching conditions and derive the corresponding Ward identities. Next, in Section \ref{sec:subsoft}, we prove the equivalence between the subleading soft theorems and the Ward identities. Finally, we explain in Section \ref{sec:magLGT} the interpretations of the charges that correspond to these new Ward identities.

\section{Asymptotic Behavior of Gauge Field}\label{sec:prelim}

\subsection{Preliminaries}

In this section, we introduce the notations employed in this paper (following the conventions of \cite{He:2019jjk}) and review related previous work.

\paragraph{Spacetime Coordinates} We work in flat null coordinates $x^\mu = (u,r,x^a)$, $a=1,\ldots,d$, where $d \geq 2$. These are related to Cartesian coordinates by
\begin{equation}
\begin{split}\label{coorddef}
	X^A = \frac{r}{2} \left( 1 + x^2 + \frac{u}{r} , 2 x^a , 1 - x^2 - \frac{u}{r} \right) ,
\end{split}
\end{equation}
and the standard Minkowski metric in flat null coordinates takes the form
\begin{align}
	ds^2 = -du\,dr + r^2\delta_{ab}dx^a\,dx^b.
\end{align}
$\ci^\pm$ is located at $r \to \pm\infty$ while keeping $(u,x^a)$ fixed, and these surfaces have the topology $S^d \times \mrr$. The point coordinatized by $x^a$ on $\ci^+$ is antipodal to the point with the same coordinate on $\ci^-$. The boundaries of $\ci^+$ and $\ci^-$ are located at $u = \pm \infty$ and are denoted by $\ci^+_\pm$ and $\ci^-_\pm$, respectively.

\paragraph{Momenta Coordinates} We will focus on the scattering of massless particles, which satisfy $p^A p_A  =0$. We parameterize such momenta using flat null coordinates so that 
\begin{equation}
\begin{split}\label{mompar}
p^A(\o,x) = \o {\hat p}^A(x) , \qquad {\hat p}^A(x) = \frac{1}{2} \left( 1 + x^2 , 2 x^a , 1 - x^2 \right) . 
\end{split}
\end{equation}
Massless gauge fields transform under the vector representation of the little group $SO(d)$ and have $d$ polarizations. The $d$ polarization vectors $\ve^A_a(x)$  are 
\begin{equation}
\begin{split}\label{polpar}
\ve^A_a(x) = \p_a {\hat p}^A(x) = \left( x_a , \delta^b_a , - x_a \right) . 
\end{split}
\end{equation}
One particle in- and out-states with momenta $\vec p$ are created from the vacuum by $in$ ($-$) and $out$ ($+$) creation and annihilation operators denoted by $\CO_\a^{(\pm)\dag}(p)$ and $\CO_\a^{(\pm)}(p)$, where $\a$ labels the polarization of the particle. They are canonically normalized, i.e.
\begin{equation}
\begin{split}
\left[ \CO^\pmm_\a(p)  , \CO^{\pmm\dagger}_\b\big(p') \right\} =  \delta_{\a\b} \big(2p^0\big) (2\pi)^{d+1} \delta^{(d+1)}\left(\vec{p} - \vec{p}\,'\right) ,
\end{split}
\end{equation}
where $[\cdot,\cdot\}$ indicates an anticommutator if both operators are fermionic and a commutator otherwise. Using the parameterization \eqref{mompar}, this can be written as
\begin{equation}
\begin{split}
\left[ \CO^\pmm_\a(\o,x) , \CO^{\pmm\dagger}_\b (\o',x') \right\} &= 2 \o^{1-d} (2\pi)^{d+1} \delta_{\a\b}  \delta\big(\o-\o'\big) \delta^{(d)}\big(x-x'\big) . 
\end{split}
\end{equation}

\paragraph{Poincar\'e Algebra} The Poincar\'e algebra is generated by $P_A$ and $M_{AB}$ and takes the form
\begin{equation}
\begin{split}
\left[P_A , P_B \right] &= 0 , \qquad \left[ P_A , M_{BC} \right] = - i ( \eta_{AB} P_C - \eta_{AC} P_B  )  \\
\left[ M_{AB} , M_{CD} \right] &= i ( \eta_{AC} M_{BD} + \eta_{BD} M_{AC} - \eta_{AD} M_{BC} - \eta_{BC} M_{AD}) . 
\end{split}
\end{equation}
We define
\begin{equation}
\begin{split}\label{Poincaregendef}
	P_\pm = - P_0 \mp P_{d+1}, \quad T_a = M_{0a} - M_{(d+1)a}, \quad D = M_{(d+1)0} , \quad K_a = M_{0a} + M_{(d+1)a} ,
\end{split}
\end{equation}
so that the nonzero commutators in the Poincar\'e algebra are given by
\begin{equation}
\begin{split}\label{Poincarealgebra}
	\left[M_{ab}, M_{cd}\right] &= i(\delta_{ac}M_{bd} + \delta_{bd}M_{ac} - \delta_{bc}M_{ad} - \delta_{ad}M_{bc}) \\
\left[ M_{ab} , T_c  \right] &=  i ( \delta_{ac} T_b - \delta_{bc} T_a )  , \qquad \left[  M_{ab} , K_c  \right] =  i ( \delta_{ac} K_b - \delta_{bc} K_a )   \\
\left[ T_a , D \right] &= i T_a , \qquad \left[ K_a , D \right] = - i K_a  , \qquad \left[ T_a , K_b\right] = - 2 i ( \delta_{ab} D + M_{ab} )   \\
\left[ M_{ab} , P_c  \right] &=  i ( \delta_{ac} P_b - \delta_{bc} P_a )  , \qquad \left[ P_\pm , D \right] = \mp i P_\pm , \qquad \left[ P_a , T_b \right] = - i \delta_{ab} P_-   \\
\left[ P_+ , T_a \right] &= - 2 i P_a ,  \qquad  \left[P_a , K_b \right] =  - i \delta_{ab} P_+  , \qquad \left[ P_- , K_a \right] = - 2 i P_a. 
\end{split}
\end{equation}
In addition to the Poincar\'e transformations, we will also  consider scale transformations $X^A \to \lambda X^A$, which appears as an effective symmetry in the infrared (near the asymptotic regions of spacetime). We denote the generator of scale transformations by $S$, which satisfies the commutation relations
\begin{equation}
\begin{split}
[ P_A , S ] = i P_A , \qquad [ M_{AB} , S ] = 0 . 
\end{split}
\end{equation}

\paragraph{Gauge Theory} A $U(1)$ gauge theory is described in terms of a 2-form field strength $F_{\mu\nu}$ that satisfies Maxwell's equations, i.e.
\begin{equation}
\begin{split}
\nabla^\mu F_{\mu\nu} = e^2 J_\nu , \qquad F_{\mu\nu} = \p_\mu A_\nu - \p_\nu A_\mu , \qquad \nabla^\mu J_\mu = 0 , 
\end{split}
\end{equation}
where $J_\mu$ is the matter current. The theory is invariant under the gauge transformations
\begin{equation}
\begin{split}
A_\mu ~\to~ A_\mu + \p_\mu \ve , \qquad \Psi_i ~\to~ e^{ i Q_i \ve } \Psi_i ,
\end{split}
\end{equation}
where $\ve \sim \ve + 2\pi$ and $Q_i \in \mathbb Z$ is the $U(1)$ charge of the matter field $\Psi_i$. Gauge transformations that vanish at infinity map physically equivalent solutions to each other and are therefore merely redundancies of the theory. We will use this redundancy to impose the gauge condition
\begin{equation}
\begin{split}\label{gaugecond}
A_u = 0
\end{split}
\end{equation}
when carrying out the asymptotic expansion of the radiative field. Note that \eqref{gaugecond} is consistent with the choice of polarization in \eqref{polpar}.

In flat null coordinates, Maxwell's equations take the form
\begin{equation}
\begin{split}\label{MaxwellEq}
	e^2 J_u &= \p_u \left( 2 \p_u A_r - \frac{1}{r^{2}} \p^a  A_a \right)  \\
	e^2J_r &= - \frac{2}{r^{d}}  \p_r \left( r^d \p_u A_r \right)  - \frac{1}{r^{2}}  \p_r \p^a A_a + \frac{1}{r^{2}} \p^2  A_r    \\
	e^2J_a &= - 2 \p_u  \p_r A_a + 2 \p_u \p_a A_r   - \frac{2}{ r^{d-2}} \p_u \p_r \left( r^{d-2} A_a \right)  - \frac{1}{r^{2}} \p_a  \p^b A_b + \frac{1}{r^{2}} \p^2 A_a ,
\end{split}
\end{equation}
where $\p^2 \equiv \p^a\p_a$. The gauge field can be split into two pieces $A_\mu = A^{(R)}_\mu  + A^{(C)}_\mu$. The radiative field $A^{(R)}_\mu$ satisfies the sourceless Maxwell's equations, whereas the Coulombic field $A^{(C)}_\mu$ is the inhomogeneous solution to Maxwell's equations and is uniquely fixed by a choice of Green's function.\footnote{We remark that in dimensions $d>2$, the Coulombic field falls off more quickly in powers of $1/r$ than the radiative field.} We are interested in the asymptotic $in$ $(-)$ and $out$ $(+)$ solutions, which are respectively obtained by choosing the retarded and advanced Green's functions. The corresponding radiative and Coulombic fields are then denoted by $A_\mu^{(R\pm)}$ and $A_\mu^{(C\pm)}$.

\subsection{Radiative Field}

In this subsection, we study the radiative gauge field near $\ci^\pm$ by expanding it in powers of $1/r$. Because we adopt the same strategy and techniques introduced in \cite{He:2019jjk}, we refer the reader there for additional details and explanations.

The radiative gauge field satisfies the sourceless Maxwell's equations and hence admits the mode expansion
\begin{equation}
\begin{split} \label{gaugefieldmodeexp}
A^{(R\pm)}_A (X) = e \int \frac{d^{d+1} q}{ ( 2\pi )^{d+1} } \frac{1}{2q^0} \left( \ve^a_{A} ({\vec q}\,) \CO^\pmm_a(\vec{q}\,) e^{ i q \cdot X }  +  \ve^a_{A} ({\vec q}\,)^* \CO^{\pmm\dagger}_a(\vec{q}\,) e^{ - i q \cdot X }  \right) ,
\end{split}
\end{equation}
where $q^0 = | \vec{q}\,|$ and $\ve^a$ is the polarization vector defined in \eqref{polpar}. Switching to the flat null coordinate parametrization of momenta defined in \eqref{mompar}, we obtain
\begin{equation}
\begin{split}\label{Aexp}
	A^{(R\pm)}_r (u,r,x) &= \frac{e}{2 ( 2 \pi )^{d+1} r  } \int_0^\infty  d\o\, \o^{d-2} \int d^d y  \left( i  \p^a  \CO^\pmm_a(\o,x+y) e^{  - \frac{i}{2} \o u - \frac{i}{2} \o r y^2 }  + \cc \right)   \\
	A^{(R\pm)}_a (u,r,x) &= \frac{er}{2 ( 2 \pi )^{d+1} } \int_0^\infty  d\o\, \o^{d-1} \int d^d y  \left(  \CO^\pmm_a(\o,x+y) e^{  - \frac{i}{2} \o u - \frac{i}{2} \o r y^2 }  + \cc \right) .
\end{split}
\end{equation}
To perform the large $|r|$ expansion, we assume that the creation and annihilation operators admit the Fourier expansion
\begin{equation}
\begin{split}\label{fourexp}
	\mathcal O_a^{(\pm)}(\w,x) = \int \frac{d^dk}{(2\pi)^d} \SO^{(\pm)}_a(\w,k) e^{ i k \cdot x },
\end{split}
\end{equation}
and that the Fourier coefficients in turn admit a soft expansion, i.e. they could be written as\footnote{In the soft expansion given, we are ignoring potential $\log\w$ terms.}
\begin{equation}
\begin{split}\label{softexp}
	\SO^\pmm_a(\o,x) = \sum_{n=0}^\infty \o^{n-1} \SO_a^{(\pm,n)}(x).
\end{split}
\end{equation}
Substituting \eqref{fourexp} and \eqref{softexp} into \eqref{Aexp}, and performing the integral over $\o$, we obtain
\begin{equation}
\begin{split}\label{Aexpfinal}
A^{(R\pm)}_r (u,r,x) &= -\frac{e}{  ( 2 \pi )^{\frac{d}{2}+1}  }  \sum_{n=0}^\infty\int \frac{d^dk}{(2\pi)^d} \left[ \frac{ i e^{ i k \cdot x }}{ (ir)^{\frac{d}{2}+\nu_n} } k^a   \SO_a^{(\pm,n)}(k)  \frac{k^{\nu_n-1} K_{\nu_n-1} \left( k z \right)}{z^{\nu_n-1} }   + \cc \right]   \\
	A^{(R\pm)}_a (u,r,x) &= -\frac{e}{  ( 2 \pi )^{\frac{d}{2}+1} }   \sum_{n=0}^\infty  \int \frac{d^dk}{(2\pi)^d} \left[\frac{ i e^{ i k \cdot x }   }{(ir)^{\frac{d}{2}+\nu_n-1} } \SO_a^{(\pm,n)}(k) \frac{k^{\nu_n} K_{\nu_n} \left( k z \right)}{z^{\nu_n} }   + \cc \right] ,
\end{split}
\end{equation}
where $K_\nu$ is the modified Bessel function of the second kind, ${k} \equiv |\vec k|$, ${z} \equiv  \frac{ \sqrt{iu} }{ \sqrt{ir} }$, and $\nu_n = \frac{d}{2}-1+n$. Because large $|r|$ corresponds to small $z$, we can expand the Bessel function about $z=0$. This asymptotic expansion for the Bessel function is qualitatively different depending on whether $\nu_n$ is an integer ($d$ even) or a half-integer ($d$ odd), so we will consider these cases separately. 

The full large $|r|$ expansion of the radiative gauge field components in all dimensions is presented in Appendix \ref{app:asexpansion} for completeness, though for our purposes we only need the large $|r|$ expansion to evaluate $\left.(1-u\p_u)F_{ra}^{(R\pm,d)}\right|_{\ci^\pm_\mp}$ (see Section \ref{sec:matching}), where we have adopted the notation $f^{(\pm,n)}$ to denote the coefficient of $|r|^{-n}$ near $r = \pm \infty$ after expanding the field $f$ in large powers of $1/|r|$.
Thus, we only need to focus on the terms in the expansion that are $O\big(1/r^d\big)$ and $O(u^0)$; the $O(u)$ terms are projected out by $1-u\p_u$, and $O(1/u)$ terms vanish at $\ci^\pm_\mp$. In even dimensions, these terms are
\begin{equation}
\begin{split}\label{deven}
	A^{(R\pm)}_r  &= \cdots + \frac{1}{r^d} \left[ - \frac{ i e  \left( \Theta(u) - \Theta(r) \right)  }{ 2(4\pi)^{\frac{d}{2}}  \G\left(\frac{d}{2}\right)}  (-\p^2)^{\frac{d}{2}-1} \p^a  \CO_a^{(\pm,1)}(x)   \right] + \cdots \\
A^{(R\pm)}_a &= \cdots + \frac{ 1 }{  r^{d-1} }  \left[  - \frac{i e  \left( \Theta(u) - \Theta(r) \right) }{  2d (4\pi)^{\frac{d}{2}}    \G\left(\frac{d}{2}\right) }  (-\p^2)^{\frac{d}{2}} \CO_a^{(\pm,1)}(x)  \right] + \cdots ,
\end{split}
\end{equation}
where we have simplified the expression by assuming 
\begin{equation}
\begin{split}\label{prop}
\CO_a^{(\pm,0)}(x) = \CO_a^{(\pm,0)\dagger}(x) , \qquad \CO_a^{(\pm,1)}(x) = - \CO_a^{(\pm,1)\dagger} (x) . 
\end{split}
\end{equation}
These assumptions are required to cancel logarithmic divergences in the asymptotic expansion in order to render the charge finite, and are discussed in greater detail in \cite{He:2019pll,He:2019jjk}. In odd dimensions, the relevant terms are
\begin{equation}
\begin{split}\label{dodd}
A^{(R\pm)}_r &=  \cdots + \frac{1}{r^d} \left[ \frac{ i e  \Gamma (d-1)  }{ 4 \pi^{d+1} (-1)^{\frac{d-1}{2}} }  \int d^d y \frac{ \p^a \CO_a^{(\pm,1)}( y )  }{ \left[(x-y)^2\right]^{ d-1 }  } \right] + \cdots \\
A^{(R\pm)}_a &= \cdots +  \frac{1}{r^{d-1} } \left[ \frac{ i e   \Gamma(d-1) }{ 4 d \pi^{d+1} (-1)^{\frac{d-1}{2}} }  \int d^d y  \frac{ (-\p^2) \CO_a^{(\pm,1)}( y ) }{ \left[(x-y)^2\right]^{d-1}  }  \right] + \cdots  .
\end{split}
\end{equation}
It follows from \eqref{deven} that in even dimensions, 
\begin{equation}
\begin{split}\label{Fraeven}
\left.(1-u\p_u )F_{ra}^{(R\pm,d)} \right|_{\ci^\pm_\mp} &=   \mp \frac{i e  }{  d (4\pi)^{\frac{d}{2}}    \G\left(\frac{d}{2}\right) } ( -\p^2)^{\frac{d}{2}-1}   \left( d \p_a \p^b -  (d-1) \delta_a^b \p^2  \right)\CO_b^{(\pm,1)}(x),  
\end{split}
\end{equation}
whereas it follows from \eqref{dodd} that in odd dimensions, 
\begin{equation}
\begin{split}\label{Fraodd}
\left.( 1 - u \p_u ) F^{(R\pm,d)}_{ra} \right|_{\ci^\pm_\mp} &= \mp \frac{   i e  \Gamma(d-1) }{ 4 d \pi^{d+1} (-1)^{\frac{d-1}{2}}  }  \left(  d \p_a \p^b  - (d-1) \delta_a^b \p^2 \right) \int d^d y  \frac{   \CO_b^{(\pm,1)}( y ) }{ \left[(x-y)^2\right]^{d-1}    }.
\end{split}
\end{equation}

\subsection{Coulombic Field}

Following the approach taken in \cite{He:2019jjk}, we know that the Coulombic gauge field $A_\mu^{(C)}$ has a large $|r|$ expansion given by 
\begin{equation}
\begin{split}
A_r^{(C\pm)} &= \sum_{n=0}^\infty \frac{ A_r^{(C\pm,d-1+n)} }{ |r|^{d-1+n} } , \qquad A_a^{(C\pm)} = \sum_{n=0}^\infty \frac{ A_a^{(C\pm,d-2+n)} }{ |r|^{d-2+n} }  . 
\end{split}
\end{equation}
The conserved current that couples to the gauge field also admits a similar expansion:
\begin{equation}
\begin{split}
J_u &=  \sum_{n=0}^\infty \frac{ J_u^{(C\pm,d+n)} }{ |r|^{d+n} } , \qquad J_a =  \sum_{n=0}^\infty \frac{ J_a^{(C\pm,d+n)} }{ |r|^{d+n} } , \qquad J_r =  \sum_{n=0}^\infty \frac{ J_r^{(C\pm,d+2+n)} }{ |r|^{d+2+n} } .
\end{split}
\end{equation}
Substituting these expressions into Maxwell's equations \eqref{MaxwellEq}, we derive various constraint equations order-by-order in large $|r|$. In particular, we have
\begin{equation}
\begin{split}\label{max1}
\p_u A_r^{(C\pm,d-1)} = 0 , \qquad  \p_u A_a^{(C\pm,d-2)}  = 0 , \qquad (d-2) \p^a A_a^{(C\pm,d-2)} + \p^2 A_r^{(C\pm,d-1)} = 0 . \\
\end{split}
\end{equation}
These equations in turn imply
\begin{equation}
\begin{split}\label{max2}
2 \p_u^2 A_r^{(C\pm,d)} &= e^2 J_u^{(\pm,d)} \\
\pm 2 d \p_u A_a^{(C\pm,d-1)}  &= e^2 J_a^{(\pm,d)} - 2 \p_u \p_a A_r^{(C\pm,d)} - \left( \delta_a^b \p^2  - \p_a \p^b \right)  A_b^{(C\pm,d-2)} . 
\end{split}
\end{equation}
The coefficient of $|r|^{-d}$ in the expansion of $F_{ra}^{(C\pm)}$ is
\begin{equation}
\begin{split}
F_{ra}^{(C\pm,d)} = \mp (d-1) A_a^{(C\pm,d-1)} - \p_a A_r^{(C\pm,d)} . 
\end{split}
\end{equation}
Acting on both sides with $\p_u^2$ and using \eqref{max1}, \eqref{max2}, we find
\begin{equation}
\begin{split}\label{mainconstraint}
\p_u^2 F_{ra}^{(C\pm,d)} = - \frac{e^2}{2d} \left[ (d-1) \p_u J_a^{(\pm,d)} + \p_a J_u^{(\pm,d)} \right] . 
\end{split}
\end{equation}

\section{Ward Identity}\label{sec:ward}

\subsection{Matching Condition}\label{sec:matching}

In \cite{He:2019pll}, it was shown that the Ward identity corresponding to the insertion of $\p^a \CO^{(\pm,1)}_a$ is associated with the antipodal matching condition\footnote{Coordinate $x^a$ on $\ci^+$ and $\ci^-$ correspond to antipodal points on the celestial sphere.}
\begin{equation}
\begin{split}\label{origmcond}
\left.\left( 1 - u \p_u \right) F_{ur}^{(+,d+1)}  \right|_{\ci^+_-} = \left.\left( 1 - u \p_u \right) F_{ur}^{(-,d+1)}  \right|_{\ci^-_+} .
\end{split}
\end{equation}
However, we now want to derive a set of $d$ independent Ward identities involving the insertion of $\CO^{(\pm,1)}_a$, which will ultimately be equivalent to the $d$ subleading soft photon theorems (one for each polarization of the soft photon). To motivate the appropriate matching condition, we begin by noting that Maxwell's equations imply
\begin{equation}
\begin{split}\label{maxcorr}
 2 F_{ur}^{(\pm,d+1)} = \pm e^2 J_r^{(\pm,d+2)} \pm \p^a F_{ra}^{(\pm,d)}  . 
\end{split}
\end{equation}
It follows that \eqref{origmcond} is equivalent to matching $\p^a F_{ra}^{(\pm,d)}$ across spatial infinity, as the current vanishes on the boundaries of null infinity. Since this matching condition gives rise to a Ward identity corresponding to inserting $\p^a \CO^{(\pm,1)}_a$, it is natural to conjecture that in order to obtain $d$ independent Ward identities corresponding to inserting $\CO_a^{(\pm,1)}$, we require the matching condition 
\begin{equation}
\begin{split}\label{matchcond}
\left.\left( 1 - u \p_u \right) F_{ra}^{(+,d)}  \right|_{\ci^+_-} = - \left.\left( 1 - u \p_u \right) F_{ra}^{(-,d)}  \right|_{\ci^-_+} .
\end{split}
\end{equation}
If we define the charge\footnote{To verify that this is indeed the charge whose Ward identity implies the subleading soft photon theorem, it must generate appropriate divergent large gauge transformations on the in- and out-states. This is verified in the next subsection.}
\begin{equation}
\begin{split}
	\CQ^\pm_Y \equiv \pm \frac{2}{e^2} \int_{\ci^\pm_\mp} d^d x\, Y^a(x)  ( 1 - u \p_u ) F_{ra}^{(\pm ,d)}, 
\end{split}
\end{equation}
where $Y^a(x)$ is a vector field on the transverse space $\mrr^d$, the matching condition \eqref{matchcond} immediately implies
\begin{equation}
\begin{split}\label{chargematch}
\CQ^+_Y = \CQ^-_Y,
\end{split}
\end{equation}
so that the charge is classically conserved.

\subsection{Soft and Hard Charges}

In the semiclassical picture, \eqref{chargematch} implies the following Ward identity for the charge $\CQ_Y$:
\begin{equation}
\begin{split}
\bra{ \text{out} } \left( \CQ_Y^+ - \CQ_Y^- \right) \ket{\text{in}} = 0.
\end{split}
\end{equation}
Analogous to our decomposition of the gauge field $A_\mu$ into a radiative field $A_\mu^{(R)}$ and a Coulombic field $A_\mu^{(C)}$, we may decompose the charge into a soft piece and a hard piece, i.e.
\begin{align}
	\CQ^{\pm}_Y = \CQ^{\pm S}_Y + \CQ^{\pm H}_Y,
\end{align} 
where
\begin{align}
\begin{split}
	\CQ^{\pm S}_Y &\equiv \pm \frac{2}{e^2} \int_{\ci^\pm_\mp} d^d x\, Y^a(x)  ( 1 - u \p_u ) F_{ra}^{(R\pm ,d)} \\
	 \CQ^{\pm H}_Y &\equiv \pm \frac{2}{e^2} \int_{\ci^\pm_\mp} d^d x\, Y^a(x)  ( 1 - u \p_u ) F_{ra}^{(C\pm ,d)}.
\end{split}
\end{align}
Thus, the Ward identity becomes
\begin{equation}
\begin{split}\label{mainwi}
\bra{ \text{out} } \left( \CQ_Y^{+S} - \CQ_Y^{-S} \right) \ket{\text{in}} = - \bra{ \text{out} } \left( \CQ_Y^{+H} - \CQ_Y^{-H} \right) \ket{\text{in}} .
\end{split}
\end{equation}

The form of the soft charge depends on the spacetime dimension. Using \eqref{Fraeven}, the soft charge in even dimensions is
\begin{equation}
\begin{split}\label{softchargeeven}
	\CQ^{\pm S}_Y &= - \frac{i  }{ e  (4\pi)^{\frac{d}{2}}    \G\left(\frac{d}{2}+1\right) }  \int d^d y\, Y^a (y)  ( -\p^2)^{\frac{d}{2}-1}   \left( d \p_a \p^b -  (d-1) \delta_a^b \p^2  \right)\CO_b^{(\pm,1)}(y)   ,
\end{split}
\end{equation}
and using \eqref{Fraodd}, the soft charge in odd dimensions is
\begin{equation}
\begin{split}\label{softchargeodd}
\CQ^{\pm S}_Y &=  \frac{   i   \Gamma(d-1) }{ 2 d e  \pi^{d+1}  }  (-1)^{\frac{d+1}{2}} \int d^d x\, Y^a (x)   \int d^d y  \frac{   \left( d \p_a \p^b  - (d-1) \delta_a^b \p^2 \right) \CO_b^{(\pm,1)}( y ) }{ \left[(x-y)^2\right]^{d-1}  }.    
\end {split}
\end{equation}
The form of the hard charge, on the other hand, is independent of dimension, and using \eqref{mainconstraint} is given by 
\begin{equation}
\begin{split}\label{hardcharge}
	\CQ^{\pm H}_Y = \frac{1}{d}\int_{\ci^\pm} du\, d^d x\, Y^a(x) \left[ (d-1) J_a^{(\pm,d)}  - u   \p_a J_u^{(\pm,d)}  \right],
\end{split}
\end{equation}
where we have assumed that there are no stable massive particles in the system so that the contribution to $\CQ_Y$ from $\ci^\pm_\pm$ vanishes. We will demonstrate in Section \ref{LROcalc} below that\footnote{Readers who are mainly interested in the final result should feel free to skip Section \ref{LROcalc}.}
\begin{equation}\label{hardonestate}
\begin{split}
\bra{\o_i,x_i} \CQ^{+H}_Y &= \frac{2i(d-1)Q_i}{\o_id} \left[  Y^a(x_i) \p_{x_i^a}    -   \frac{1}{d-1} \p_a Y^a(x_i)  \w_i \p_{\o_i}   + i \p^a Y^b(x_i) \CS_{i\,ab} \right] \bra{\o_i,x_i}  \\
	\CQ^{-H}_Y  \ket{\o_i,x_i}   &= - \frac{2i(d-1)Q_i}{\w_id} \left[ Y^a(x_i)  \p_{x_i^a}    -    \frac{1}{d-1}\p_a Y^a(x_i)   \o_i \p_{\o_i}  + i  \p^a Y^b(x_i)  \CS_{i\,ab} \right] \ket{\o_i,x_i}.
\end{split}
\end{equation}
In an $S$-matrix element, the hard charge acts on multi-particle states as a tensor product of one-particle states. Thus, we can rewrite \eqref{mainwi} as
\begin{align}\label{mainwi2}
\begin{split}
	&\la\text{out}|\left(\CQ_Y^{+S} - \CQ_Y^{-S}\right)|\text{in}\ra  \\
	&~~ = -\sum_{i=1}^n \frac{2i(d-1)Q_i}{\w_i d}\left(Y^a(x_i)\p_{x_i^a} - \frac{1}{d-1}\p_aY^a(x_i)\w_i\p_{\w_i} + i\p^aY^b(x_i)\CS_{i\,ab}\right)\la\text{out}|\text{in}\ra.
\end{split}
\end{align}

\subsubsection{Action of Hard Charges}\label{LROcalc}

In this sub-subsection, we will prove \eqref{hardonestate}. For notational simplicity, we do not distinguish between in- and out-states and drop the superscripts $\pmm$ on all operators. We will also drop the subscripts on the annilation operators and simply denote them as $\CO(\w,x)$.

Begin by defining the light-ray operators (LROs)
\begin{equation}
\begin{split}\label{LROdef}
\mqq(x) \equiv \int_{-\infty}^\infty du\, J_u^{(d)}(u,x) , \qquad \mjj_a(x) \equiv \int_{-\infty}^\infty du \left[ (d-1) J_a^{(d)}  - u   \p_a J_u^{(d)}  \right]. 
\end{split}
\end{equation}
Note that $\mqq(x)$ is the LRO appearing in the leading hard charge (see \cite{He:2019jjk}), and $\mjj_a(x)$ is the LRO appearing in the subleading hard charge:
\begin{align}\label{LROcharge}
	Q_\ve^H = \int d^dx\, \ve(x)\mqq(x), \qquad \CQ_Y^H = \frac{1}{d}\int d^dx\,Y^a(x)\mjj_a(x).
\end{align}
Recall that massless one-particle states are created out of the vacuum by creation and annihilation operators. Since the LROs annihilate the vacuum \cite{Kravchuk:2018htv}, we have
\begin{equation}
\begin{split}\label{jjaction}
\bra{ \o , x } \mqq(x') = \bra{0} \left[ \CO(\o,x) , \mqq(x') \right]  , \qquad \bra{ \o , x } \mjj_a(x') = \bra{0} \left[ \CO(\o,x) , \mjj_a(x') \right] .
\end{split}
\end{equation}
Our objective now is to determine the following commutators
\begin{equation}
\begin{split}\label{commtoderive}
\left[ \CO(\o,x) , \mqq(x') \right] , \qquad \left[ \CO(\o,x) , \mjj_a(x') \right] . 
\end{split}
\end{equation}

We will follow the spirit of the procedure outlined in \cite{Cordova:2018ygx}, where the authors determined the commutators of LROs by using only a few basic assumptions.
Their argument relied on properties of unitary conformal field theories and crucially required invariance under special conformal transformations. In our case, we are interested in QFTs with an $S$-matrix; hence, we do not generically have access to scale or special conformal invariance. However, in theories with only massless particles, $S$-matrices have an effective scale symmetry in the low energy limit. We will utilize this effective scale symmetry (along with Poincar\'e invariance) to completely fix the commutator between the LROs in \eqref{LROdef} and the annihilation operators $\CO(\o,x)$, and is based on the following assumptions:
\begin{enumerate}

\item\label{micro} \textit{Microcausality: Spacelike separated LROs commute with each other.}

To utilize this assumption, we note that $\CO(\o,x)$ is itself a LRO. For instance, the scalar annihilation operator can be constructed out of a scalar field as
\begin{equation}
\begin{split}
\CO(\o,x) = \frac{(2\pi i)^{\frac{d}{2}}}{\w^{\frac{d}{2}-1}}  \int_{-\infty}^\infty du \,e^{\frac{i}{2} \w u } \Phi^{\left(d/2\right)} (u,x)  ,
\end{split}
\end{equation}
where $\Phi^{(d/2)} (u,x)   = \lim\limits_{r\to\infty} \left[ r^{\frac{d}{2}} \Phi(u,r,x) \right]$. This implies that $\CO(\o,x)$ is a LRO localized on the null-ray $x$ on $\ci^\pm$. This assumption then implies that the commutators \eqref{commtoderive} depend only on the Dirac delta function $\delta^{(d)}\big(x-x'\big)$ and its derivatives. 

\item\label{uni} \textit{Unitarity: All operators transform in representations of the Poincar\'e and scaling symmetry algebra.}

\item\label{wi} \textit{Ward Identities: The charge ${\hat Q} = \displaystyle\int d^dx\, \mqq(x)$ generates global $U(1)$ transformations on the operators.} 

This implies 
\begin{equation}
\begin{split}\label{globalcharge}
\left[ \CO (\o,x) , {\hat Q} \right] = Q \CO (\o,x)  .
\end{split}
\end{equation}

\item\label{mc} \textit{Minimal Coupling: The commutator of an annihilation operator with a LRO is also an annihilation operator (or derivatives thereof) of the same particle type.}

This is required in order to fix the commutators uniquely. As was shown in \cite{Elvang:2016qvq}, the subleading soft theorem receives corrections from higher-derivative operators and is therefore not universal. In this paper, we will only focus on the universal part of the subleading soft theorem in minimally coupled theories.\footnote{The non-universal part of the subleading soft theorem also has an interpretation as a Ward identity \cite{Laddha:2017vfh}.}

\end{enumerate}

Having outlined our assumptions, we now proceed to prove \eqref{hardonestate}. Begin by labeling pertinent operators by their boost charge $J$ (eigenvalue under $D$) and twist $\tau = J-\Delta$ ($\Delta$ is the eigenvalue under the scale charge $S$), which we have for convenience listed in the table below (see Appendix \ref{app:poincare} for details). We will use it to fix the commutators $\left[\CO(\w,x),\mqq(x')\right]$ and $\left[\CO(\w,x),\mjj_a(x')\right]$.
\begin{table}[h!]
  \begin{center}
  \begin{tabular}{c|c}
      Quantity & $(J,\tau)$ \\
      \hline
      $\mqq$ & $(d,d)$ \\
      $\mjj_a$ & $(d,d+1)$ \\
      $\CO_\a$ & $\left(0,\frac{d}{2}\right)$ \\
      $\p_a$ & $(1,1)$ \\
      $\w$ & $(1,0)$  
    \end{tabular}
  \end{center}
  \caption{This table tabulates the boost charges and twists of various relevant operators.}
\label{Jtab}
\end{table}
\subsection*{$\left[\CO(\o,x),\mqq(x')\right]$}

The Lorentzian separation between the two operators is $\big(x-x'\big)^2$. They are therefore spacelike separated as long as $x\neq x'$, which by Assumption  \ref{micro} implies 
\begin{equation}\label{genOQ}
\begin{split}
\left[ \CO(\o,x) , \mqq(x') \right] = \delta^{(d)}\big(x-x'\big) L(\o,x) + \p_a \delta^{(d)}\big( x - x' \big) L^a ( \o , x ) + \cdots  ,
\end{split}
\end{equation}
where $\cdots$ represents terms involving additional derivatives of the delta function. Integrating over the transverse directions $x'$, we obtain
\begin{equation}
\begin{split}
\left[ \CO(\o,x) , {\hat Q} \right]  = L(\o,x) .
\end{split}
\end{equation}
Comparing to \eqref{globalcharge}, it follows that $L(\o,x) = Q \CO(\o,x)$. 

Next, consider $L^a(\o,x)$. Utilizing Table~\ref{Jtab}, we observe that the twist of the left-hand-side of \eqref{genOQ} is $\frac{d}{2}+d$, and the twist of $\p_A \delta^{(d)}(x-x')$ is $d+1$. This implies that the twist of $L^a$ is $\frac{d}{2}-1$. On the other hand, by Assumption~\ref{mc}, $L^a$ is locally constructed from $\CO$ and must therefore have the general form\footnote{Lorentz invariance forbids explicit factors of $x^a$.}
\begin{equation}
\begin{split}
L^a(\o,x) = \sum_{n,q=0}^\infty \sum_{p=-\infty}^\infty c_{p,q}^{a a_1 \cdots a_n} \o^p  (\o\p_\o)^q \p_{a_1} \cdots \p_{a_n}  \CO(\o,x) .
\end{split}
\end{equation}
All the operators on the right-hand-side have twist $\frac{d}{2} + n$. Since $n \geq 0$, there are no operators with twist $\frac{d}{2}-1$,\footnote{The assumption of minimal coupling implies that the Poincar\'e invariant constants $c_{p,q}^{a_1 \cdots a_n}$ are dimensionless and therefore have vanishing twists.} thereby implying that $L^a(\o,x) = 0$. Likewise, terms with additional derivatives of the delta function in \eqref{genOQ} are excluded by the same argument, and so 
\begin{equation}
\begin{split}
	\left[ \CO(\o,x) , \mqq(x') \right] =  Q \delta^{(d)}\big(x-x'\big) \CO(\o,x)  . 
\end{split}
\end{equation}

\subsection*{$\left[\CO(\o,x),\mjj_a(x')\right]$}

By Assumption \ref{micro}, the commutator takes the form
\begin{equation}\label{OJgen}
\begin{split}
\left[ \CO(\o,x) , \mjj_a(x') \right] &=  \delta^{(d)}\big( x - x' \big) K^\0_a  ( \o , x )  + \p^b \delta^{(d)}\big( x - x' \big) K^\1_{ab} ( \o , x )  ,
\end{split}
\end{equation}
where we have excluded terms involving higher derivatives on the delta function using the same twist argument as above. By Assumption \ref{mc}, $K^{(0)}$ and $K^{(1)}$ have the general form
\begin{equation}\label{Kgenform}
\begin{split}
K_a^\0 (\o,x) &= \sum_{n,q=0}^\infty \sum_{p=-\infty}^\infty \big(c^\0_{p,q}\big)_a{}^{a_1 \cdots a_n} \o^p  (\o\p_\o)^q \p_{a_1} \cdots \p_{a_n}  \CO(\o,x)    \\
K_{ab}^\1 (\o,x) &= \sum_{n,q=0}^\infty \sum_{p=-\infty}^\infty \big(c^\1_{p,q}\big)_{ab}{}^{a_1 \cdots a_n} \o^p  (\o\p_\o)^q \p_{a_1} \cdots \p_{a_n}  \CO(\o,x) .
\end{split}
\end{equation}
From \eqref{OJgen}, we can deduce that the twists of $K^\0$ and $K^\1$ are $\frac{d}{2}+1$ and $\frac{d}{2}$, respectively, and Table~\ref{Jtab} further implies that the boost charges of $K^\0$ and $K^\1$ are $0$ and $-1$, respectively. 
Matching the twists and boost charges on both sides of \eqref{Kgenform}, we find that $p=-1$ for both $K^{(0)}$ and $K^{(1)}$, while $n=1$ for $K^\0$ and $n=0$ for $K^\1$. There are no constraints on $q$, so
\begin{equation}
\begin{split}
K_a^\0 (\o,x) &= \frac{1}{\o} \CR_{ab} (\o \p_\o) \p^b  \CO(\o,x)  , \qquad K_{ab}^\1 (\o,x) = \frac{1}{\o} \CB_{ab}(\o\p_\o) \CO(\o,x) ,
\end{split}
\end{equation}
where
\begin{equation}
\begin{split}
\CR_{ab}(\w\p_\w) = \sum_{q=0}^\infty    \big(c_q^\0\big)_{ab}(\w\p_\w)^q, \qquad \CB_{ab}(\w\p_\w) =  \sum_{q=0}^\infty  \big(c_q^\1 \big)_{ab}(\w\p_\w)^q . 
\end{split}
\end{equation}

As we shall next show, $\CR_{ab}$ and $\CB_{ab}$ can be fixed by  checking consistency with the Jacobi identity
\begin{equation}
\begin{split}\label{JacobiIdentity}
\left[ \left[ \CO(\o,x) , \mjj_a(x') \right]  , X \right] = \left[ \CO(\o,x) ,  \left[ \mjj_a(x') , X \right] \right]  + \left[ \left[ \CO(\o,x) , X \right]  , \mjj_a(x') \right]   ,
\end{split}
\end{equation}
where $X$ is a generator of the Poincar\'e algebra. In particular, choosing $X$ to be $P_-$, $P_a$, $K_a$, and $M_{ab}$ suffices to fix $\CR_{ab}$ and $\CB_{ab}$. Readers interested in detailed derivations of the commutators below should refer to Appendix \ref{app:poincare}.

\begin{itemize}
\item $P_-$: We note
\begin{equation}
\begin{split}
\left[ \CO(\o,x) , P_- \right] &= - \o \CO(\o,x) , \qquad \left[ \mjj_a(x) , P_- \right] = - 2 i \p_a \mqq(x) . 
\end{split}
\end{equation}
\eqref{JacobiIdentity} then implies
\begin{equation}
\begin{split}
\left[ \CR_{ab}  ( \o \p_\o ) ,  \o \right] = 0 , \qquad  \left[ \CB_{ab} (\o\p_\o) , \o \right] + 2 i Q \delta_{ab}\w = 0 ,
\end{split}
\end{equation}
which means
\begin{equation}\label{genform}
\begin{split}
\CR_{ab}  ( \o \p_\o )  = \CR'_{ab} , \qquad \CB_{ab} (\o\p_\o) = - 2 i Q \delta_{ab} \o \p_\o + \CB'_{ab} ,
\end{split}
\end{equation}
where $\CR'_{ab}$ and $\CB'_{ab}$ are operators independent of $\w\p_\w$.

\item $P_a$: We note
\begin{equation}
\begin{split}
\left[ \CO(\o,x) , P_a  \right] &= - \o x_a \CO(\o,x) , \qquad \left[ \mjj_a(x) , P_b \right] = - 2 i x_b \p_a \mqq(x) - 2 i d \delta_{ab} \mqq(x) . 
\end{split}
\end{equation}
Using this, \eqref{JacobiIdentity} and \eqref{genform} imply
\begin{equation}\label{Rab}
\begin{split}
	\CR_{ab}(\w\p_\w) = \CR'_{ab} = 2  i  Q( d - 1 ) \delta_{ab} . 
\end{split}
\end{equation}

\item $M_{ab}$: We note
\begin{equation}
\begin{split}
\left[ \CO(\o,x) , M_{ab} \right] &= -i ( x_a\p_b -x_b\p_a )  \CO (\o,x) + \CS_{ab} \CO(\o,x)  \\
\left[ \mjj_a(x)   , M_{bc}\right] &= -i ( x_b\p_c -x_c\p_b ) \mjj_a(x) - i \delta_{ab}\mjj_c(x) + i \delta_{ac} \mjj_b(x) . 
\end{split}
\end{equation}
Using this, the Jacobi identity implies
\begin{equation}
\begin{split}\label{SBcomm}
 \left[ \CS_{ab} ,  \CB'_{cd} \right]   = i\left( \delta_{ac} \CB'_{bd} + \delta_{bd} \CB'_{ac} - \delta_{bc}  \CB'_{ad}  - \delta_{ad} \CB'_{bc} \right),
\end{split}
\end{equation}
which is simply the statement that $\CB'_{ab}$ is an $SO(d)$ covariant matrix.

\item $K_a$: We note
\begin{equation}
\begin{split}
 \left[ \CO(\o,x), K_a \right] &= i \left( x^2\p_a -  2 x_a x^b \p_b + 2x_a\o \p_{\o} \right) \CO(\o,x) + 2x^b \CS_{ab} \CO(\o,x)  \\
\left[ \mjj_a(x)   , K_b \right] &=  i \left[    x^2 \p_b - 2 x_b \left( x^c \p_c  + d   \right) \right] \mjj_a(x) +  2i  x_a \mjj_b(x)   - 2 i \delta_{ab}   x^c \mjj_c(x) . \\
\end{split}
\end{equation}
\eqref{JacobiIdentity} along with \eqref{Rab} and \eqref{SBcomm} then implies
\begin{equation}
\begin{split}\label{eq349}
	2 Q ( d-1) \CS_{ab} =  \delta_{ab} \delta^{cd} \CB'_{cd} - \CB'_{ba} .
\end{split}
\end{equation}
Symmetrizing and antisymmetrizing the indices $ab$ on both sides yields
\begin{align}
\CB'_{(ab)} = \delta_{ab}\delta^{cd}\CB'_{cd}, \qquad \CB'_{[ab]} = 2Q(d-1)\CS_{ab}.
\end{align}
Tracing \eqref{eq349} over $ab$, we immediately find $\delta^{ab} \CB'_{ab} = 0 \implies \CB'_{(ab)} = 0$. Hence, it follows by \eqref{genform} that
\begin{align}\label{Bab}
	\CB_{ab}(\w \p_\w) = -2iQ\delta_{ab}\w\p_\w + 2Q(d-1)\CS_{ab}.
\end{align}
\end{itemize}
Having determined both $\CR_{ab}$ in \eqref{Rab} and $\CB_{ab}$ in \eqref{Bab}, we substitute everything back into \eqref{OJgen} to get
\begin{equation}
\begin{split}\label{finalans}
	\left[ \CO(\o,x) , \mjj_a(x') \right] &=   \frac{2iQ}{\w} \left[  (d-1) \delta^{(d)}\big( x-x'  \big) \p_a    -    \p_a \delta^{(d)}\big( x-x' \big)  \o \p_\w \right. \\
&\left. \qquad \qquad \qquad \qquad \qquad \qquad \quad - i  (d-1) \p^b \delta^{(d)}\big( x - x' \big) \CS_{ab} \right] \CO (\o,x) . 
\end{split}
\end{equation}
Substituting this into \eqref{jjaction} and noting \eqref{LROcharge}, we obtain \eqref{hardonestate}, as promised.

\section{Connection to the Subleading Soft Theorem}\label{sec:subsoft}

We begin by recalling the subleading soft photon theorem in its standard momentum coordinates. Let $\CA_n(p_1, \cdots ,p_n)$ be a scattering amplitude involving $n$ particles with momenta $p_1, \ldots ,p_n$, and let $\CA_{n+1}^\text{out} (\vec{p}_\g,\ve_a ; p_1 , \cdots , p_n )$ be the same amplitude with an additional outgoing photon with momentum $p_\g$ and polarization $\ve_a$. In a minimally coupled theory, the soft limit ($E_\g \to 0$) of the amplitude has the form
\begin{equation}
\begin{split}\label{outsoftth}
 \CA^{\text{out}}_{n+1} ( p_\g,\ve_a ; p_1 , \cdots , p_n ) &= O \left( \frac{1}{E_\g} \right)  + S_a^\1 \CA_n ( p_1 , \cdots , p_n  ) + O \left( E_\g \right) , 
\end{split}
\end{equation}
where the $\frac{1}{E_\g}$ pole is related to the leading soft photon theorem \cite{Weinberg:1995mt,Weinberg:1965nx}, and 
\begin{equation}
\begin{split}\label{softfactors}
S_a^\1 = - i e \sum_{i=1}^n Q_i \frac{ p_\g^A \ve_a^B (\vec p_\g )  }{ p_i \cdot p_\g } \CJ_{i\,AB}  
\end{split}
\end{equation}
is the subleading soft factor. Here, $\CJ_{i\,AB}$ is the angular momentum operator, which is the sum of the orbital and spin angular momenta:
\begin{align}
\begin{split}
	\CJ_{i\,AB} = \CL_{i\,AB} + \CS_{i\,AB} = - i \left( p_{i A} \pd{}{p_i^B} - p_{iB} \pd{}{p_i^A} \right)  + \mathcal S_{i\,AB} . 
\end{split}
\end{align}

To write the subleading soft factor in flat null coordinates, we parametrize $p_i^A$ and $p_\g^A$ via \eqref{mompar} as $(\w_i,x_i^a)$ and $(\w,x^a)$, respectively. Using \eqref{polpar}, we have
\begin{equation}
\begin{split}
S_a^\1 = e \sum_{i=1}^n \frac{Q_i}{\o_i} \left[ \p^b \log \left[ ( x - x_i )^2 \right]  \left( \delta_{ab} \o_i \p_{\o_i} - i \CS_{i\,ab} \right) - \CI_a{}^b( x - x_i ) \p_{x_i^b} \right],
\end{split}
\end{equation}
where
\begin{equation}\label{invariant}
\begin{split}
	\CI_{ab}(x) = \frac{x^2}{2}\p_a\p_b\log x^2 = \delta_{ab} - \frac{2 x_a x_b }{x^2 }  . 
\end{split}
\end{equation}
By the LSZ reduction formula, the left-hand-side of \eqref{outsoftth} corresponds to the insertion of the operator $\CO_a^{(+,1)}(x) -  \CO_a^{(-,1)}(x)$ in the $S$-matrix.\footnote{See Appendix C of \cite{He:2019jjk} for details.} One can then rewrite the subleading soft photon theorem as
\begin{equation}
\begin{split}\label{subsoft}
&\bra{\text{out}} \left( \CO_b^{(+,1)}(x)  - \CO_b^{(-,1)}(x)   \right) \ket{ \text{in} } \\
&\qquad \quad =  e \sum_{i=1}^n \frac{Q_i}{\o_i} \left[ \p^c \log \left[ ( x - x_i )^2 \right]  \left( \delta_{bc} \o_i \p_{\o_i} - i \CS_{i\,bc} \right) - \CI_b{}^c( x - x_i ) \p_{x_i^c} \right] \braket{ \text{out} }{ \text{in} }.
\end{split}
\end{equation}

\subsection{Soft Theorem $\implies$ Ward Identity}\label{sec:softtoward}

In this subsection, we want to use the subleading soft photon theorem \eqref{subsoft} to derive the Ward identity \eqref{mainwi2}. For the even dimensional case, we act on both sides of \eqref{subsoft} with the operator
\begin{align}
	- \frac{i  }{    e  (4\pi)^{\frac{d}{2}}    \G\left(\frac{d}{2}+1\right) }  \int d^d x\, Y^a (x)  ( -\p^2)^{\frac{d}{2}-1}   \left( d \p_a \p^b -  (d-1) \delta_a^b \p^2  \right).
\end{align}	
Applying \eqref{softchargeeven} and using the fact that in even dimensions
\begin{align}\label{even1}
	(-\p^2)^{\frac{d}{2}}\log\left[(x-x_i)^2\right] = -(4\pi)^{\frac{d}{2}}\G\left(\frac{d}{2}\right)\delta^{(d)}(x-x_i),
\end{align}
we obtain
\begin{align}
\begin{split}
	&\la\text{out}|\left(\CQ_Y^{+S} - \CQ_Y^{-S}\right)|\text{in}\ra = \\
	&~~~~ = -\sum_{i=1}^n \frac{2i(d-1)Q_i}{\w_i d}\left(Y^a(x_i)\p_{x_i^a} - \frac{1}{d-1}\p_aY^a(x_i)\w_i\p_{\w_i} + i\p^aY^b(x_i)\CS_{i\,ab}\right)\la\text{out}|\text{in}\ra,
\end{split}
\end{align}
which is precisely \eqref{mainwi2}.

For the odd dimensional case, we act on both sides of \eqref{subsoft} with the operator
\begin{align}
	\frac{   i   \Gamma(d-1) }{ 2 d e  \pi^{d+1}  }  (-1)^{\frac{d+1}{2}} \int d^d y\, Y^a (y)   \int d^d x  \frac{  d \p_a \p^b  - (d-1) \delta_a^b \p^2 }{ \left[(x-y)^2\right]^{d-1}  }.
\end{align}
Applying \eqref{softchargeodd} and  using the fact that in odd dimensions
\begin{align}\label{odd1}
	\int d^dx \frac{\p^2\log\left[(x-x_i)^2\right]}{\left[(x-y)^2\right]^{d-1}} = \frac{4(-1)^{\frac{d-1}{2}}\pi^{d+1}}{\G(d-1)}\delta^{(d)}(y-x_i),
\end{align}
we again obtain \eqref{mainwi2}.

\subsection{Ward Identity $\implies$ Soft Theorem}

Now, we want to prove the converse, that the Ward identity \eqref{mainwi2} implies the subleading soft theorem \eqref{subsoft}, thereby proving that the two are completely equivalent. First, recall \eqref{invariant} and choose $Y^a(z) = \CY^a(z) \equiv \CI^a{}_b(x-z) \zeta^b$ in \eqref{mainwi2} for a constant $\zeta$, so that
\begin{equation}
\begin{split}
\left[  d  \p^a \p_b  - (d-1) \delta^a_b  \p^2   \right] \CY^b (z) = (d-1) \zeta^a (-\p^2) \log \left[ ( x - z )^2 \right],
\end{split}
\end{equation}
where the derivatives are with respect to $z$. Substituting this choice of $Y^a$ into \eqref{softchargeeven} and \eqref{softchargeodd}, and using \eqref{even1} and \eqref{odd1}, we determine in both even and odd dimensions that the soft charge takes the form\footnote{In proving this, we have integrated by parts in \eqref{softchargeodd} and neglected potential boundary terms. To justify this, we can take $Y^a = \CY^a$ in a region encompassing all the points $x_i$ in the amplitude and zero outside, in which case all boundary terms are trivially vanishing.}
\begin{equation}
\begin{split}\label{eq354}
\CQ^{\pm S}_\CY = \frac{ 2 i }{ d e }  (d-1) \zeta^a \CO_a^{(\pm,1)}(x) .
\end{split}
\end{equation}
Substituting this into \eqref{mainwi2} with $Y^a(z) = \CY^a(z)$ and simplifying yields
\begin{align}
\begin{split}
	& \zeta^a\la\text{out}|\left(\CO_a^{(+,1)}(x) - \CO_a^{(-,1)}(x)\right)|\text{in}\ra  \\
	&~~~~ = e\sum_{i=1}^n \frac{Q_i}{\w_i }\zeta^a\left[\p^c\log\left[(x-x_i)^2\right]\left(\delta_{ac}\w_i\p_{\w_i} - i\CS_{i\,ac}\right) -\CI_a{}^c(x-x_i)\p_{x_i^c}  \right]\la\text{out}|\text{in}\ra.
\end{split}
\end{align}
Choosing $\zeta^a = \delta^a_b$ yields
\begin{equation}
\begin{split}
&\bra{\text{out}} \left( \CO_b^{(+,1)}(x)  - \CO_b^{(-,1)}(x)   \right) \ket{ \text{in} } \\
&\qquad \quad =  e \sum_{i=1}^n \frac{Q_i}{\o_i} \left[ \p^c \log \left[ ( x - x_i )^2 \right]  \left( \delta_{bc} \o_i \p_{\o_i} - i \CS_{i\,bc} \right) - \CI_b{}^c( x - x_i ) \p_{x_i^c} \right] \braket{ \text{out} }{ \text{in} },
\end{split}
\end{equation}
which is precisely the subleading soft photon theorem given in \eqref{subsoft}.

\section{Electric and Magnetic Large Gauge Transformations} \label{sec:magLGT}

In previous sections, we have established the equivalence between the subleading soft theorem and a Ward identity for the charge
\begin{equation}
\begin{split}
\CQ^\pm_Y \equiv  \pm \frac{2}{e^2} \int_{\ci^\pm_\mp} d^d x \, Y^a  \left( 1 - u \p_u \right) F_{ra}^{(\pm,d)} .
\end{split}
\end{equation}
In this section, we will show that this charge generates divergent electric \textit{and} magnetic large gauge transformations. We begin by using Hodge decomposition to decompose the vector field as 
\begin{equation}
\begin{split}
Y^a(x) = \p^a \lambda(x) + \p_b K^{ab} (x) , \qquad K^{ab} (x) = - K^{ba} (x) . 
\end{split}
\end{equation}
We can then write the charge as\footnote{We assume $\lambda$ and $K^{ab}$ fall off sufficiently quickly in $|x|$ so we can integrate by parts and neglect boundary terms.}
\begin{equation}
\begin{split}\label{QYterm}
\CQ^\pm_Y \equiv \mp \frac{2}{e^2} \int_{\ci^\pm_\mp} d^d x \, \lambda\left( 1 - u \p_u \right)    \p^a F_{ra}^{(\pm,d)}  \mp \frac{4}{e^2} \int_{\ci^\pm_\mp} d^d x  K^{ab}  \left( 1 - u \p_u \right) \p_{[a} F_{b]r}^{(\pm,d)}  . 
\end{split}
\end{equation}
Using \eqref{maxcorr} and the fact that the current vanishes at $\ci^\pm_\pm$, we can simplify the first term as
\begin{equation}
\begin{split}
\mp \frac{2}{e^2} \int_{\ci^\pm_\mp} d^d x\, \lambda \left( 1 - u \p_u \right)    \p^a F_{ra}^{(\pm,d)} = - \frac{4}{e^2} \int_{\ci^\pm_\mp} d^d x\, \lambda \left( 1 - u \p_u \right)    F_{ur}^{(\pm,d+1)} .
\end{split}
\end{equation}
This is precisely the charge studied in \cite{He:2019pll}, where it was shown to generate divergent \emph{electric} large gauge transformations. 

To simplify the second term in \eqref{QYterm}, we use the Bianchi identity
\begin{equation}
\begin{split}
\p_r F_{ab} + 2\p_{[a}F_{b]r} = 0   \quad \implies \quad  2\p_{[a} F_{b]r}^{(\pm,d)} = \pm (d-1) F_{ab}^{(\pm,d-1)}  . 
\end{split}
\end{equation}
It then follows that the second term in \eqref{QYterm} reduces to
\begin{equation}\label{secondterm}
\begin{split}
 \mp \frac{4}{e^2} \int_{\ci^\pm_\mp} d^d x  \, K^{ab} \left( 1 - u \p_u \right)  \p_{[a} F_{b]r}^{(\pm,d)} =  - \frac{2}{e^2} (d-1)\int_{\ci^\pm_\mp} d^d x \, K^{ab}    \left( 1 - u \p_u \right)  F_{ab}^{(\pm,d-1)} ,
\end{split}
\end{equation}
which, as we shall now show, generates divergent \emph{magnetic} large gauge transformations. 

In form notation, the charge that generates magnetic gauge transformations on a hypersurface $\Sigma$ is\footnote{In form notation, the charge that generates electric gauge transformations is $Q^\Sigma_\ve = \frac{1}{e^2} \displaystyle\int_\Sigma \ve \ast F$. }
\begin{equation}\label{maghodge}
\begin{split}
{\tilde Q}_\CK^\Sigma = \frac{1}{2\pi} \int_{\p\Sigma} \CK \wedge F ,
\end{split}
\end{equation}
where $\CK$ is a $(d-2)$-form. This acts on the dual gauge field via
\begin{equation}
\begin{split}
{\tilde A}  \to {\tilde A} + d \CK ,
\end{split}
\end{equation}
where $d {\tilde A} = \ast d A$. To study this charge with $\Sigma=\ci^\pm$, we note
\begin{equation}
\begin{split}
( \CK \wedge F )_{a_1 \cdots a_d } &= \frac{d(d-1)}{2}\CK_{[a_1\cdots a_{d-2}} F_{a_{d-1} a_d] } =  \frac{1}{2} \e_{a_1 \cdots a_d}   (\ast_d \CK )^{ab} F_{ab}, 
\end{split}
\end{equation}
where $\ast_d$ is the Hodge dual on $\mrr^d$. Substituting this into \eqref{maghodge} with $\Sigma = \ci^\pm$ yields
\begin{equation}
\begin{split}
{\tilde Q}_\CK^{\ci^\pm} = \pm  \frac{1}{4\pi}  \int_{\ci^\pm_\mp}  d^d x \,\lim_{r \to \pm\infty} \left[ |r|^d  (\ast_d \CK )^{ab} F_{ab} \right] .
\end{split}
\end{equation}
To match the charges, we take
\begin{equation}
\begin{split}
(\ast_d \CK)^{ab} = -   \frac{8\pi}{r e^2 } ( d - 1 ) K^{ab}.
\end{split}
\end{equation}
Recalling that the field strength near $\ci^\pm_\mp$ admits the expansion
\begin{equation}
\begin{split}
F_{ab} (u,r,x) = \frac{F_{ab}^{(\pm,d-2)}(u,x) }{ |r|^{d-2} } +  \frac{F_{ab}^{(\pm,d-1)}(u,x) }{ |r|^{d-1} } + \cdots, 
\end{split}
\end{equation}
the charge simplifies to
\begin{equation}\label{divergentQ}
\begin{split}
{\tilde Q}_\CK^{\ci^\pm} = -  \frac{2}{ e^2 }  ( d - 1 )   \int_{\ci^\pm_\mp}  d^d x\,K^{ab}\left( |r|  F_{ab}^{(\pm,d-2)} +F_{ab}^{(\pm,d-1)} \right).
\end{split}
\end{equation}
This charge is formally divergent, and it is in this sense that the symmetry generated by this charge is a divergent magnetic gauge symmetry. However, when this divergent charge is inserted into an $S$-matrix element, the divergent contribution (the terms that are $O(|r|)$, which are shown explicitly above, as well as terms that are $O(u)$, which are implicitly present in $F_{ab}^{(\pm,d-1)}$) vanishes due to the constraint equation
\begin{equation}
\begin{split}
\bra{ \text{out} } \left(\p_a \CO_b^{(\pm,0)} (x)  - \p_b \CO_a^{(\pm,0)} (x)\right)\ket{\text{in}}  = 0 ,
\end{split}
\end{equation}
thereby giving rise to a finite Ward identity. Comparing \eqref{divergentQ} with \eqref{secondterm}, we see that the finite part of the charge is exactly \eqref{secondterm}, the second term of \eqref{QYterm}. This concludes our demonstration that the asymptotic symmetry dual to the subleading soft photon theorem is a divergent electric and magnetic large gauge symmetry.

\section*{Acknowledgements}

We would like to thank Clay Cordova, Daniel Kapec, Monica Pate, Ana-Maria Raclariu and Shu-Heng Shao for useful conversations. TH is grateful to be supported by U.S. Department of Energy grant DE-SC0009999 and by funds from the University of California. PM gratefully acknowledges support from U.S. Department of Energy grant DE-SC0009988.

\appendix

\section{Asymptotic Expansions}\label{app:asexpansion}

In this appendix, we list the full large $|r|$ expansion of the radiative gauge field components $A_r$ and $A_a$ near $\ci^\pm$. Recall that \eqref{Aexpfinal} states
\begin{equation}
\begin{split}\label{Aexpfinalapp}
A^{(R\pm)}_r (u,r,x) &= -\frac{e}{  ( 2 \pi )^{\frac{d}{2}+1}  }  \sum_{n=0}^\infty\int \frac{d^dk}{(2\pi)^d} \left[ \frac{ i e^{ i k \cdot x }}{ (ir)^{\frac{d}{2}+\nu_n} } k^a   \SO_a^{(\pm,n)}(k)  \frac{k^{\nu_n-1} K_{\nu_n-1} \left( k z \right)}{z^{\nu_n-1} }   + \cc \right]   \\
	A^{(R\pm)}_a (u,r,x) &= -\frac{e}{  ( 2 \pi )^{\frac{d}{2}+1} }   \sum_{n=0}^\infty  \int \frac{d^dk}{(2\pi)^d} \left[\frac{ i e^{ i k \cdot x }   }{(ir)^{\frac{d}{2}+\nu_n-1} } \SO_a^{(\pm,n)}(k) \frac{k^{\nu_n} K_{\nu_n} \left( k z \right)}{z^{\nu_n} }   + \cc \right] ,
\end{split}
\end{equation}
where $K_\nu$ is the modified Bessel function of the second kind, ${k} \equiv |\vec k|$, ${z} \equiv  \frac{ \sqrt{iu} }{ \sqrt{ir} }$, $\nu_n = \frac{d}{2}-1+n$, and $\SO_a^{(\pm)}$ are the Fourier coefficients of the annihilation operator $\CO_a^{(\pm)}$ (see \eqref{fourexp} and \eqref{softexp}). Because large $|r|$ is equivalent to small $z$, we can perform a large $|r|$ expansion of these components by utilizing the known expansions about $z=0$ for the Bessel functions and then performing an inverse Fourier transform. For further details, we refer the reader to \cite{He:2019jjk}, where this procedure was introduced and more carefully explained.

In even dimensions $D=d+2>4$, we only need the $z=0$ expansion of Bessel functions with nonnegative integer orders. Carrying out the procedure outlined above, we obtain \small{
\begin{equation}
\begin{split}\label{devenapp}
	A^{(R\pm)}_r  =& - \frac{e}{ 16\pi^{\frac{d}{2}+1}  }  \sum_{n=0}^\infty \sum_{s=0}^{\nu_n-2} \frac{ ( - 1 )^s \G(\nu_n-s-1)   }{ 2^{2s-n} \G(s+1) }  \frac{ (iu)^{s+1-\nu_n} }{ (ir)^{\frac{d}{2}+s+1} }  ( -\p^2)^s  \p^a  \CO_a^{(\pm,n)}(x) \\
&~ - \frac{8e}{ (4\pi)^{\frac{d}{2}+1} }  \sum_{n=0}^\infty   \sum_{s=0}^\infty \frac{ (-1)^{\nu_n} \left(  \log \frac{\sqrt{iu}}{\sqrt{ir}}  + c_{s,\nu_n-1}  \right)  }{ 2^{2s+n} \G(s+1) \G(s+\nu_n) }    \frac{  (iu)^s (-\p^2)^{s+\nu_n-1} \p^a  \CO_a^{(\pm,n)}(x)  }{ (ir)^{d-1+n+s} }   \\
&~  - \frac{ 2^d e}{ 16  \pi   }  \sum_{n=0}^\infty  \sum_{s=0}^\infty \frac{ \G \left(d-2+n+s\right) }{   2^{-n} (-1)^s \G(s+1)   }  \frac{   (iu)^s }{ (ir)^{d-1+n+s} }  \int d^d y  \frac{ \p^a \CO^{(\pm,n)}_a ( y ) }{  \left[(x-y)^2\right]^{d-2+n+s} }  + \cc \\
	A^{(R\pm)}_a =& - \frac{e}{ 8 \pi^{\frac{d}{2} +1} }   \sum_{n=0}^\infty \sum_{s=0}^{\nu_n-1}  \frac{ ( - 1 )^s \G(\nu_n-s)    }{ 2^{2s-n} \G(s+1) } \frac{ i (iu)^{s-\nu_n}  }{(ir)^{\frac{d}{2}+s-1} }   ( -\p^2)^s  \CO_a^{(\pm,n)}(x)   \\
&~ - \frac{e}{  2^d \pi^{\frac{d}{2} +1} }   \sum_{n=0}^\infty  \sum_{s=0}^\infty \frac{ (-1)^{\nu_n-1} \left(   \log \frac{\sqrt{iu}}{\sqrt{ir}}  + c_{s,\nu_n}  \right)  }{ 2^{2s+n} \G(s+1) \G(s+\nu_n+1) }  \frac{ i (iu)^s (-\p^2)^{s+\nu_n} \CO_a^{(\pm,n)}(x) }{(ir)^{d-2+n+s} }     \\
&~  - \frac{ 2^d e}{ 8 \pi  }   \sum_{n=0}^\infty  \sum_{s=0}^\infty \frac{   \G \left(d-1+n+s\right)  }{ 2^{-n} (-1)^s\G(s+1)  } \frac{ i (iu)^s }{(ir)^{d-2+n+s} }   \int d^d y  \frac{ \CO^{(\pm,n)}_a ( y ) }{  \left[(x-y)^2\right]^{d-1+n+s} } + \cc,
\end{split}
\end{equation}}
where $c_{s,\nu} \equiv \g_E - \log 2 - \frac{1}{2}(H_{s} + H_{s +\nu})$, $\g_E$ the Euler-Mascheroni constant, and $H_n = \sum_{k=1}^n \frac{1}{k}$.

For $D=4$, in addition to needing the $z=0$ expansion of Bessel functions with nonnegative integer orders, we also need the $z=0$ expansion of $K_{-1}(kz)$. $A_a^{(R\pm)}$ stays the same as \eqref{devenapp} with $d=2$, but the final expression for $A_r^{(R\pm)}$ is \small{
\begin{equation}
\begin{split}\label{deven2}
A^{(R\pm)}_r  =&     \frac{e}{ 8\pi^3  } \frac{1}{ir}  \int d^2 y  \frac{  ( x - y )^a }{( x - y )^2} \CO_a^{(\pm,0)}(y) + \frac{e}{ 4 \pi  } \sum_{s=0}^\infty \frac{  (-1)^s\G (s+1)  }{ \G(s+2) }  \frac{  (iu)^{s+1} }{ (i r)^{s+2}  } \int d^2 y  \frac{  \p^a \CO^{(\pm,0)}_a ( y ) }{  \left[(x-y)^2\right]^{s+1} } \\
	&~- \frac{e}{ 8 \pi^2  }  \sum_{s=0}^\infty \frac{  \log \frac{\sqrt{iu}}{\sqrt{ir}}   + c_{s,1}   }{ 2^{2s} \G(s+1) \G(s+2) }  \frac{  (iu)^{s+1}  }{ (i r)^{s+2} } ( -\p^2)^s \p^a \CO_a^{(\pm,0)}(x) \\
	&~  - \frac{e}{ 16\pi^2  }  \sum_{n=1}^\infty \sum_{s=0}^{n-2} \frac{ ( - 1 )^s \G(n-s-1)   }{ 2^{2s-n} \G(s+1) }  \frac{ (iu)^{s-n+1} }{ (ir)^{s+2} }  ( -\p^2)^s  \p^a  \CO_a^{(\pm,n)}(x) \\
	&~  - \frac{e}{ 2 \pi^2 }  \sum_{n=1}^\infty   \sum_{s=0}^\infty \frac{ (-1)^{n} \left(  \log \frac{\sqrt{iu}}{\sqrt{ir}}  + c_{s,n-1}  \right)  }{ 2^{2s+n} \G(s+1) \G(n+s) }    \frac{  (iu)^s  }{ (ir)^{n+s+1} } (-\p^2)^{n+s-1} \p^a  \CO_a^{(\pm,n)}(x)   \\
	&~  - \frac{e}{4\pi}  \sum_{n=1}^\infty  \sum_{s=0}^\infty \frac{2^n \G (n+s ) }{(-1)^s \G(s+1)   }  \frac{   (iu)^s }{ (ir)^{n+s+1} }  \int d^2 y  \frac{ \p^a \CO^{(\pm,n)}_a ( y ) }{ \left[(x-y)^2\right]^{n+s} }  + \cc.
\end{split}
\end{equation}}

Finally, in odd dimensions $D=d+2>4$, we only need the $z=0$ expansion of Bessel functions with nonnegative half-integer orders. Repeating the above procedure yields \small{
\begin{equation}
\begin{split}\label{doddapp}
A^{(R\pm)}_r =& - \frac{e}{  16 \pi^{\frac{d}{2}}  }  \sum_{n=0}^\infty  \sum_{s=0}^\infty    \frac{  2^{n-2s} (-1)^{\frac{d-1}{2} +n }   }{ \G(s+1)\G ( s - \nu_n +2) }  \frac{(iu)^{s-\nu_n+1}  }{ (ir)^{\frac{d}{2}+1+s} } (-\p^2)^s \p^a \CO_a^{(\pm,n)}(x) \\
	&~  - \frac{e}{ 16 \pi^{d+1}  }  \sum_{n=0}^\infty   \sum_{s=0}^\infty  \frac{ \Gamma (d-2+n+s)}{ 2^{-n} (-1)^s \Gamma (s+1)} \frac{ (iu)^s }{ (ir)^{d-1+n+s} }  \int d^d y \frac{ \p^a \CO_a^{(\pm,n)}( y )  }{ \left[(x-y)^2\right]^{ d-2+n+s }  }  + \cc \\
	A^{(R\pm)}_a =&   \frac{e}{  8\pi^{\frac{d}{2}} }   \sum_{n=0}^\infty \sum_{s=0}^\infty    \frac{ 2^{n-2s} (-1)^{\frac{d-1}{2}+n}    }{ \G(s+1)\G ( 1 + s - \nu_n ) }  \frac{ i (iu)^{s-\nu_n} }{(ir)^{\frac{d}{2}-1+s} } [-\p^2]^s \CO_a^{(\pm,n)}(x) \\
	&~  - \frac{e}{  8 \pi^{d+1} }   \sum_{n=0}^\infty    \sum_{s=0}^\infty   \frac{\Gamma(d-1+n+s)}{   2^{-n} (-1)^s  \Gamma (s+1)} \frac{ i (iu)^s }{(ir)^{d-2+n+s} } \int d^d y \frac{ \CO_a^{(\pm,n)}( y )  }{ \left[(x-y)^2\right]^{ d-1+n+s }  } + \cc .
\end{split}
\end{equation}}

\section{Poincar\'e and Scale Transformations}\label{app:poincare}

In this appendix, we determine the Poincar\'e and scale transformations of the annihilation operators and the LROs \eqref{LROdef}. The goal is to derive Table \ref{Jtab} and the commutators used when verifying the Jacobi identity \eqref{JacobiIdentity}. For simplicity, both subscripts labeling the annihilation operators and superscripts $\pm$ labeling the currents and LROs will be kept implicit.

We first derive the necessary commutators involving annihilation operators, which transform as
\begin{equation}
\begin{split}
\left[ \CO ( p ) , P_A  \right] &= - p_A \CO ( p ) \\
\left[ \CO ( p ) , M_{AB}  \right] &=  - i   \left( p_A \p_{p^B} - p_B \p_{p^A}  \right) \CO(p)  +    \CS_{AB} \CO(p) \\
\left[ \CO (p) , S \right] &= - i\left( p^A \p_{p^A}  + \frac{d}{2} \right) \CO(p)  . 
\end{split}
\end{equation}
Parametrizing the momenta in flat null coordinates, i.e.
\begin{equation}
\begin{split}\label{appmompar}
p^A(\o,x) = \o {\hat p}^A(x) , \qquad {\hat p}^A(x) = \frac{1}{2} \left( 1 + x^2 , 2 x^a , 1 - x^2 \right),
\end{split}
\end{equation}
and recalling that we defined in \eqref{Poincaregendef}
\begin{equation}
\begin{split}\label{appPoincaregendef}
	P_\pm = - P_0 \mp P_{d+1}, \quad T_a = M_{0a} - M_{(d+1)a}, \quad D = M_{(d+1)0} , \quad K_a = M_{0a} + M_{(d+1)a},
\end{split}
\end{equation}
we find
\begin{equation}
\begin{split}
	\left[ \CO(\o,x) , P_- \right] &=  - \o \CO(\o,x) , \qquad \left[ \CO(\o,x) , P_+ \right] = -  \o x^2 \CO(\o,x)  \\
	\left[ \CO(\o,x) , P_a \right] &=  - \o x_a \CO(\o,x), \qquad \left[\CO(\o,x), T_a\right] = i\p_a \CO(\o,x)  \\
	\left[\CO(\o,x), M_{ab}\right] &= -i ( x_a\p_b -x_b\p_a )  \CO(\o,x) + \CS_{ab} \CO(\o,x) \\
	\left[\CO(\o,x), D \right] &= i ( x^a\p_a -\o \p_{\o} ) \CO(\o,x) \\
	\left[\CO(\o,x), K_a \right] &= i \left( x^2\p_a -  2 x_a x^b \p_b + 2x_a\o \p_{\o} \right) \CO(\o,x) + 2x^b \CS_{ab} \CO (\o,x)  \\
	\left[ \CO (\o,x) ,S  \right] &= i\left( - \o \p_\o - \frac{d}{2} \right) \CO(\o,x).
\end{split}
\end{equation}

Next, we want to determine the Poincar\'e and scale transformations of the LROs defined in \eqref{LROdef}, which we repeat here for convenience:
\begin{equation}
\begin{split}\label{appLROdef}
\mqq(x) \equiv \int_{-\infty}^\infty du\, J_u^{(d)}(u,x) , \qquad \mjj_a(x) \equiv \int_{-\infty}^\infty du \left[ (d-1) J_a^{(d)}  - u   \p_a J_u^{(d)}  \right]. 
\end{split}
\end{equation}
Observing that a conserved current obeys the commutators 
\begin{equation}
\begin{split}
\left[ J_A(X)  , P_B \right] &= i \p_B J_A(X) \\
\left[ J_A(X) , M_{BC} \right] &= - i \left( X_B \p_C - X_C \p_B \right) J_A(X) - i   \eta_{AB} J_C(X) + i \eta_{AC} J_B(X) \\
\left[ J_A(X) , S \right] &= i \left( X^B \p_B + d + 1 \right) J_A(X),
\end{split}
\end{equation}
we have in flat null coordinates the $J_u$ commutators 
\begin{equation}
\begin{split}\label{Jutransform}
	[ J_u , P_- ] &= - 2  i \p_u J_u, \qquad [ J_u , P_+ ] = - 2 i \left( x^2 \p_u + \p_r - \frac{1}{r}  x^a   \p_a  \right) J_u  \\
	[ J_u , P_a ] &= - 2 i \left( x_a \p_u - \frac{1}{2r} \p_a \right) J_u, \qquad [J_u, T_a] = i\p_a J_u  \\
	[J_u, M_{ab}] &= -i ( x_a\p_b -x_b\p_a )  J_u \\
	 [J_u, D] &= i ( x^a\p_a  + u \p_u - r \p_r  + 1  ) J_u  \\
	[J_u, K_a ] &=  i \left[  \left( x^2  + \frac{u}{r}   \right) \p_a - 2 x_a  \left( x^b \p_b  + u \p_u - r \p_r  + 1 \right) \right] J_u  + \frac{i}{r} J_a  \\
	[ J_u , S ] &= i (  u \p_u + r \p_r + d + 1 ) J_u ,
\end{split}
\end{equation}
and the $J_a$ commutators
\begin{equation}
\begin{split}\label{Jatransform}
	[ J_a , P_- ] &=  - 2 i  \p_u J_a, \qquad [ J_a , P_+ ] =  - 2i  \left[ x^2 \p_u + \p_r - \frac{1}{r}  \left( x^b \p_b  + 1 \right) \right] J_a - 4 i  x_a J_u \\
	[ J_a , P_b ] &= - 2 i  \left( x_b \p_u - \frac{1}{2r} \p_b \right) J_a - 2 i  \delta_{ab} J_u , \qquad [J_a, T_b ] = i\p_b J_a  \\
	[J_a, M_{bc}] &= -i ( x_b\p_c -x_c\p_b )  J_a - i \delta_{ab} J_c + i \delta_{ac} J_b  \\
	[J_a, D ] &= i \left( x^b\p_b  + u \p_u - r \p_r  + 1  \right) J_a  \\
	[J_a, K_b ] &=  i \left[   \left( x^2 + \frac{u}{r} \right) \p_b - 2 x_b \left( x^c \p_c + u \p_u - r \p_r  + 1 \right) \right]  J_a  \\
&\qquad \qquad \qquad \qquad \qquad \qquad \qquad  +  2i  x_a  J_b    - 2 i \delta_{ab}  \left( x^c J_c + u J_u  -  r  J_r \right) \\
	[ J_a , S ]  &= i (  u \p_u + r \p_r + d ) J_a .
\end{split}
\end{equation}
Extracting the $1/|r|^d$ coefficient from $J_u$ and $J_r$, and then taking the limit $r \to \pm\infty$, we obtain
\begin{equation}
\begin{split}\label{Judtransform}
	\big[ J^\d_u , P_- \big] &=  - 2 i \p_u J^\d_u  , \qquad \big[ J^\d_u , P_+ \big] = - 2 i x^2 \p_u J^\d_u  \\
	\qquad \big[ J^\d_u , P_a \big] &= - 2 i x_a \p_u J^\d_u, \qquad \big[J^\d_u, T_a\big] = i\p_a J^\d_u  \\
	\big[J^\d_u, M_{ab}\big] &= -i ( x_a\p_b -x_b\p_a )  J^\d_u \\
	\big[J^\d_u, D \big] &= i ( x^a\p_a  + u \p_u + d + 1  ) J^\d_u  \\
	\big[J^\d_u, K_a \big] &=  i \left[  x^2  \p_a - 2 x_a  \left( x^b \p_b  + u \p_u  + d  + 1 \right) \right] J^\d_u \\
	\big[ J^\d_u , S  \big] &= i \left[  u \p_u  + 1 \right] J^\d_u 
\end{split}
\end{equation}
and
\begin{equation}
\begin{split}
	\big[ J^\d_a , P_- \big] &= - 2 i  \p_u J^\d_a  , \qquad \big[ J^\d_a , P_+ \big] = - 2 i x^2 \p_u J^\d_a - 4 i x_a J^\d_u \\
	\big[ J^\d_a , P_b \big] &= - 2 i x_b \p_u J^\d_a - 2 i  \delta_{ab} J^\d_u , \qquad \big[J^\d_a, T_b \big] = i\p_b J^\d_a  \\
	\big[J^\d_a, M_{bc}\big] &= -i ( x_b\p_c -x_c\p_b )  J^\d_a - i \delta_{ab} J^\d_c + i \delta_{ac} J^\d_b  \\
	\big[J^\d_a, D \big] &= i \left( x^b\p_b  + u \p_u + d   + 1  \right) J^\d_a  \\
	\big[J^\d_a, K_b \big] &=  i \left[    x^2 \p_b - 2 x_b \left( x^c \p_c + u \p_u + d   + 1 \right) \right]  J^\d_a  \\
&\qquad \qquad \qquad \qquad \qquad  +  2i  x_a  J^\d_b    - 2 i \delta_{ab}  \left( x^c J^\d_c + u J^\d_u \right)\\
	\big[ J_a^\d , S \big]  &= i u \p_u J^\d_a.
\end{split}
\end{equation}

Finally, we integrate over $u$ to determine the commutators involving the LROs given in \eqref{appLROdef}. For $\mqq(x)$, we have
\begin{equation}
\begin{split}\label{Qtransform}
	[ \mqq(x) , P_- ] &=0 , \qquad [ \mqq(x) , P_+ ] = 0 \\
	[ \mqq(x) , P_a ] &= 0 , \qquad [\mqq(x), T_a] = i\p_a \mqq(x)  \\
	[\mqq(x), M_{ab}] &= -i ( x_a\p_b -x_b\p_a )  \mqq(x) \\
	[\mqq(x), D ] &= i ( x^a\p_a  + d) \mqq(x)  \\
	[\mqq(x), K_a ] &=  i \left[  x^2  \p_a - 2 x_a  \big( x^b \p_b    + d    \big) \right] \mqq(x) \\
	[\mqq(x), S ] &= 0,
\end{split}
\end{equation}
and for $\mjj_a(x)$, we have
\begin{equation}
\begin{split}\label{Jtransform}
[ \mjj_a(x)   , P_- ] &= - 2 i\p_a  \mqq(x)  , \qquad [ \mjj_a(x)   , P_+ ] = - 2  i  x^2 \p_a   \mqq(x)  - 4 i d x_a \mqq(x) \\
[\mjj_a(x)   , P_b ] &= - 2 i x_b \p_a   \mqq(x)  - 2 i d \delta_{ab}  \mqq(x)  , \qquad [ \mjj_a(x)   , T_b ] = i  \p_b \mjj_a(x) \\
[ \mjj_a(x)   , M_{bc}] &= -i ( x_b\p_c -x_c\p_b ) \mjj_a(x) - i \delta_{ab}\mjj_c(x) + i \delta_{ac} \mjj_b(x)  \\
[ \mjj_a(x)   , D ] &= i \left( x^b\p_b + d   \right) \mjj_a(x)  \\
[ \mjj_a(x)   , K_b ] &=  i \left[    x^2 \p_b - 2 x_b \left( x^c \p_c  + d   \right) \right] \mjj_a(x) +  2i  x_a \mjj_b(x)   - 2 i \delta_{ab}   x^c \mjj_c(x) \\
[ \mjj_a(x)   , S ] &=  -i \mjj_a(x).
\end{split}
\end{equation}

\bibliography{HM-bib}{}
\bibliographystyle{utphys}

\end{document}